\documentclass{PoS}
\usepackage[authoryear]{natbib}
\bibpunct{(}{)}{;}{a}{}{,}

\title{The Physics of the Cold Neutral Medium: Low-frequency Carbon Radio Recombination
Lines with the Square Kilometre Array}

\ShortTitle{Low-frequency RRLs with the SKA}

\author{

\speaker{ J.~B.~R.~Oonk}$^{1,2}$, L.~K.~Morabito$^1$, F.~Salgado$^1$, M.~C.~Toribio$^2$, R.~J.~van Weeren$^3$, A.~G.~G.~M.~Tielens$^1$, H.~J.~A.~R\"ottgering$^1$
\\
$^{1}$Leiden Observatory, Leiden University, PO Box 9513, NL-2300 RA Leiden, the Netherlands\\
$^{2}$Netherlands Institute for Radio Astronomy (ASTRON), Postbus 2, NL-7990 AA Dwingeloo, the Netherlands\\
$^{3}$Harvard-Smithsonian Center for Astrophysics, 60 Garden Street, Cambridge, MA 02138, USA
\\
E-mail: \email{oonk@astron.nl}
}

\abstract{
The Square Kilometre Array (SKA) will transform our understanding of the role of the cold, atomic gas in galaxy evolution. The interstellar medium (ISM) is the repository of stellar ejecta and the birthsite of new stars and, hence, a key factor in the evolution of galaxies over cosmic time. Cold, diffuse, atomic clouds are a key component of the ISM, but so far this phase has been difficult to study, because its main tracer, the HI 21~cm line, does not constrain the basic physical information of the gas (e.g., temperature, density) well.

The SKA opens up the opportunity to study this component of the ISM through a complementary tracer in the form of low-frequency ($<$350 MHz) carbon radio recombination lines (CRRL). These CRRLs provide a sensitive probe of the physical conditions in cold, diffuse clouds. The superb sensitivity, large field of view, frequency resolution and coverage of the SKA allows for efficient surveys of the sky, that will revolutionize the field of low-frequency recombination line studies. 

By observing these lines with the SKA we will be able determine the thermal balance, chemical enrichment, and ionization rate of the cold, atomic medium from degree-scales down to scales corresponding to individual clouds and filaments in our Galaxy, the Magellanic Clouds and beyond. Furthermore, being sensitive only to the cold, atomic gas, observations of low-frequency CRRLs with the SKA will aid in disentangling the warm and cold constituents of the HI 21~cm emission.
}

\FullConference{
Advancing Astrophysics with the Square Kilometre Array\\
June 8-13, 2014\\
Giardini Naxos, Italy}

\newcommand{\skipthis}[1]{}

\newcommand\apj{ApJ}
\newcommand\mnras{MNRAS}
\newcommand\aanda{A\&A}
\newcommand\nat{Nature}

\begin{document}

\section{The cold neutral medium} 
The ISM is central to the evolution of galaxies as the birth site of new stars and the repository of processed stellar ejecta. Star formation slowly consumes matter from the ISM and later returns it to the ISM via winds and explosions, enriched by the nucleosynthetic products of stellar interiors. Over cosmic time, this constant recycling and enrichment drives galaxy evolution. Stars also control the radiative energy budget of the ISM and its emission characteristics. Photons from massive stars with energies above 13.6~eV ionize hydrogen creating HII regions. Less energetic photons couple to the gas through photoelectrons from large molecules and small dust grains that heat atomic gas in PhotoDissociation regions (PDRs) surrounding HII regions and -- on a much larger scale -- in diffuse interstellar clouds. Through their winds and explosions, stars also stir up the ISM dynamically, sweeping up gas and forming large bubbles. This injection of mechanical energy into the ISM is a source of turbulent pressure, supporting the gas disk and the clouds therein against galactic- and self-gravity. Radiative and mechanical interaction of stars with their environment results in a number of distinct phases in the ISM; e.g., a phase of cold, neutral atomic hydrogen -- diffuse clouds or Cold Neutral Medium (CNM) -- and a phase of warm, neutral atomic hydrogen -- the Warm Neutral Medium (WNM). 

Clearly, there is a complex feedback between stars and the ISM that determines the structure, composition, chemical evolution, and observational characteristics of the ISM in the Milky Way and other galaxies all the way back to the first galaxies. Understanding this interaction has to start with characterizing the phases of the ISM and linking these to stellar and galactic energy sources and physical processes that couple interstellar gas to the mechanical and radiative energy inputs from stars and the large scale motions in the galaxy.

Over the last decades, our understanding of the ISM and its role in the evolution of galaxies has increased dramatically, but in leaps, driven, in particular, by the opening up of new observational windows by unique observatories. While the molecular medium has been well characterized and its role in star formation is well documented, analyzing and interpreting the atomic phase from the HI 21~cm line data in terms of the density and temperature has been challenging and has sofar only be done in a very limited number of cases (e.g. Heiles \& Troland 2003). The main difficulty here is to accurately separate the warm and cold components that both contribute to the observed HI 21~cm signal. A consequence of this is that the thermal balance of the cold, atomic ISM and its role in starformation is poorly understood. 

A promising, complementary method is to observe low-frequency CRRLs. These CRRLs have been found to be a good tracer of cold atomic gas (e.g. Payne et al. 1989, Anantharamaiah et al. 1994, Kantharia \& Anantharamaiah 2001, Anish-Roshi et al. 2001). Their observed properties, i.e. line strength and width, as a function of frequency provides a convenient method to determine the physical conditions of this gas. The unprecedented sensitivity and bandwidth of the SKA will revolutionize the field of low-frequency, i.e. $<350~MHz$, CRRLs. In Sect. 2 we give a brief overview of the current status of the field. In Sect. 3 we discuss the models necessary to interprete CRRL observations. In Sect. 4 we describe how SKA-LOW will impact and transform the field of low-frequency CRRL studies. We focus our discussion here on two prime targets, the Galactic center and the Magellanic Clouds. In Sect.'s 5 and 6 we discuss the capabilities of SKA-LOW in phase 1 and 2 respectively, and finally we present a brief summary in Sect.~7.

\section{Low-frequency CRRLs}
Radio recombination lines (RRL) result from atoms recombining with electrons in an ionised plasma and can broadly be separated in to two classes: discrete and diffuse (e.g. see Gordon \& Sorochenko 2009 for a recent overview). Discrete RRLs from hydrogen (HRRL), helium (HeRRL) and carbon (CRRL) are typically observed above about 1~GHz. Here the RRL spectrum is dominated by HRRL and HeRRL lines that are found to be excellent tracers of the dense, warm, ionized gas (e.g. Palmer \& Zuckerman 1966). Discrete RRLs are described in more detail by M. Thompson et al. (2014) and R. Beswick et al. (2014) in this volume.

In this chapter we focus on the lower-frequency, i.e. $<$350 MHz, diffuse RRLs. At these frequencies we predominantly observe RRLs from carbon and hydrogen (e.g. Konovalenko \& Sodin 1981, Anantharamaiah 1985.). HeRRLs have not been detected at these frequencies. Recent overviews of low-frequency RRLs and the instruments that are able to observe them are given in Peters et al. (2011) and Gordon \& Sorochenko (2009). Below we will briefly summarize the current status of low-frequency RRL observations.

\subsection{Galactic CRRLs}
\noindent \textbf{Large-scale studies}: A summary of CRRL observations below 200~MHz is presented in Peters et al. (2011). Two large-scale Galactic surveys below 200~MHz were done at 76 MHz (Erickson et al. 1995) and 34.5 MHz (Kantharia \& Anantharamaiah 2001). These low-resolution (beam FWHM $\ge$2~deg) studies find that CRRL absorption is widespread throughout the inner Galactic plane with typical peak optical depths in the range 0.5$\times$10$^{-3}$ to 1$\times$10$^{-3}$ and typical line widths in the range 10 to 50 km/s. HRRLs were not detected by these surveys. 

Two other low resolution (beam FWHM $\sim$2~deg) surveys were conducted near 330 MHz by Anantharamaiah (1985) and Roshi et al. (2000). These surveys also find widespread RRL emission, from hydrogen and carbon, in the inner Galactic plane with peak optical depths in the range -0.5$\times$10$^{-3}$ to -2$\times$10$^{-3}$ for HRRL and -0.5$\times$10$^{-3}$ to -1$\times$10$^{-3}$ for CRRL (the minus sign indicates emission). The CRRLs are found to have typical linewidths in the range 15-30 km/s, about a factor 2 narrower than the corresponding HRRLs with typical linewidths 30-60 km/s.

The broader HRRLs likely arise in the warm ionized medium (WIM) (e.g. Anantharamaiah 1985, Heiles et al. 1996, Anish-Roshi et al. 2001, Kantharia et al. 2007). The CRRLs are more likely associated with the CNM as shown by their good correspondence with HI 21~cm absorption (e.g. Anish-Roshi et al. 2001, Anantharamaiah et al. 1994, Oonk et al. 2014). Theoretical models of CRRL emission are consistent with this picture (see Sect. 3).

Kantharia \& Anantharamaiah (2001) made an attempt at modelling the large-scale CRRL emission by combining the above mentioned surveys. They found that the observed CRRLs are consistent with arising in the CNM. However, their analysis was limited by the low resolution, poor spatial overlap and narrow frequency sampling of these surveys. Especially the unknown beam dilution due to the low resolution affected their investigation. Higher spatial resolution observations are therefore essential to make further progress in this field.
\\
\\
\noindent \textbf{Small-scale studies}: The only source that has been studied with sufficient spatial resolution, spectral coverage and sensitivity to allow for detailed CRRL modeling are the cold clouds along the line of sight towards Cassiopeia~A (Cas~A). The CRRLs towards this source can reach a peak optical depth of almost 10$^{-2}$ (e.g. Kantharia et al. 1998). This is a factor of a few brighter than seen on large scales in other regions, although this could simply a consequence of the low spatial resolution used in the studies of these other regions. The CRRL line width observed towards Cas~A varies from about 4~km/s at 350~MHz to about 10~km/s at 50~MHz (Payne et al. 1989; Oonk et al. in prep.). 

Modeling shows that the CRRLs observed towards Cas~A arise in the CNM  (e.g. Payne et al. 1989; Kantharia et al. 1998). Cas~A is an example of a bright Galactic source that can be used for detailed high-resolution CRRL 'pinhole' studies of the CNM. Similarly extragalactic sources can also be used to perform such pinhole studies, as shown by our recent detection of CRRL absorption from the Milky Way against the bright extragalactic source Cygnus~A (Oonk et al. 2014). 

With LOFAR we are conducting a Galactic survey to establish the physical scales on which CRRLs are present. Some preliminary results are shown in Fig.~1. These and previous observations show that CRRLs are widespread in our Galaxy with typical peak optical depths of a few times 10$^{-4}$ to a few times 10$^{-3}$ and typical line widths from a few to a few tens of km/s.

\subsection{Extragalactic CRRL studies}
The first detection of extragalactic low-frequency CRRLs has recently been achieved with LOFAR. In Morabito et al. (2014) we report CRRL absorption from M82 near 60 MHz with a peak optical depth of about 3$\times$10$^{-3}$ and we find a good correspondence with HI 21~cm absorption. This first detection shows the potential that with increasing sensitivity and bandwidth it is now becoming possible to detect CRRLs in other galaxies. In the upcoming years we will use LOFAR to carry out a flux limited, i.e. F$_{\nu}>$10~Jy at 160 MHz, survey for extragalactic CRRLs. The primary aim of this survey will be to establish what type and fraction of bright extragalactic radio sources show detectable CRRL signals and how the CRRLs compare to other properties of these systems.

\section{CRRL models \& methods}
Low-frequency CRRLs originate in the partially ionized ($X_e\simeq 10^{-4}$), dense atomic phases of diffuse clouds and PDRs. This can be understood as singly ionized carbon has an ionization potential below that of hydrogen and this is therefore the dominant ionization state of carbon in the ISM. The observed properties of CRRLs, i.e. line strength and width, provide a convenient probe of the physical conditions of the gas. However, low-frequency CRRLs are not in local thermal equilibrium (LTE), and therefore non-LTE models are necessary to analyze the observations in terms of electron temperature, electron density and carbon column density  (e.g. Shaver 1975; Brocklehurst \& Salem 1977; Walmsley \& Watson 1982; Ponomarev \& Sorochenko 1992; Storey \& Hummer 1995; Salgado et al. in prep.). A comprehensive literature exists on the topic of non-LTE models for low-frequency RRL emission and for further details we refer the interested reader to the above references and references therein. Here we show a few examples of such models in Fig.~2 and we briefly discuss how these models can be used to obtain the physical conditions of the CRRL emitting gas.

\textit{Electron temperature and density}: The optical depth of CRRLs is affected by departure coefficients from LTE. For a given temperature T$_{e}$ and density n$_{e}$ these coefficients trace out a unique curve as function of frequency (or equivalently quantum number n), see Fig.~2. Therefore by measuring the CRRLs over a range of frequencies and comparing with models we can obtain T$_{e}$ and n$_{e}$ (e.g. Kantharia et al. 1998). Additional constraints on both of these parameters can be obtained by measuring how the pressure broadening affects the line width of the CRRLs (e.g. Kantharia et al. 1998; Oonk et al. 2014).

\textit{Thermal pressure}: Once the electron temperature and density of the CRRL emitting gas has been obtained one can estimate the associated thermal pressure by invoking a carbon to hydrogen abundance [C/H]. The thermal pressure is then given by p$_{th}\sim$n$_{e}$$\times$T$_{e}$/[C/H] (e.g. Payne et al. 1989, Kantharia et al. 1998).

\textit{Carbon abundance}: A systematic uncertainty in deriving hydrogen densities from CRRLs and thus thermal pressures is related to the carbon abundance [C/H]. Typically [C/H] has an uncertainty of about a factor 2 in our Galaxy. However, [C/H] can be determined by comparing, for the same line of sight, the carbon column density obtained from CRRLs with the cold hydrogen column density obtained from HI~21 cm absorption studies.

\textit{WNM to CNM fraction}: Fig.~2 shows that low-frequency CRRLs only trace cold, atomic gas, i.e. the CNM. This means that CRRLs can be used as a tracer of this phase in terms of temperature and distribution. To also obtain the hydrogen density it is necessary to invoke a carbon to hydrogen abundance [C/H]. A Galactic [C/H] grid can be created by comparing CRRL and HI 21~cm absorption measurements (e.g. this work; McClure-Griffiths et al. 2014 in this volume). If [C/H] does not vary strongly between the HI 21~cm absorption measurements then the CRRLs can be used to also determine the CNM hydrogen density between absorption measurements.
 
\textit{Hydrogen ionization}: Hydrogen in diffuse clouds can only be ionized through cosmic rays. Therefore the HRRL to CRRL ratio, if they measure the same gas, is a measure of the cosmic ray ionization rate (e.g. Sorochenko \& Smirnov 1987). Recently, using observations from the WSRT at 350~MHz and LOFAR at 220~MHz, we have found a weak, narrow HRRL component in the cold clouds towards Cas~A. These HRRLs have the same width and velocity as the corresponding CRRLs indicating that the arise from the same gas and thus can be used to determine the hydrogen ionization rate, see Fig~3. 

\section{Square Kilometre Array}
The properties of the low-frequency CRRLs discussed in Sect.~2 \& 3 show that these originate in the CNM where they are associated with cold, diffuse clouds and PDRs. Here we propose to use SKA-LOW to perform a survey of Carbon Radio Recombination Lines (CRRLs) originating from cold atomic gas in diffuse interstellar clouds and in PDRs associated with regions of massive star formation in the Milky Way as well as with AGN and starburst activity in other galaxies. This CRRL survey can help answer a number of important open questions related to the ISM, such as;\\

\noindent \textit{(1) What is the morphology of CRRL emitting regions and how do they relate with the neutral, molecular, star forming and hot gas, as well as the large-scale structure of the Milky Way? \\ \\ (2) What is the thermal, pressure balance for cold, diffuse clouds? \\ \\ (3) What is the C abundance in the diffuse gas and how does this vary with galactic radius? \\ \\ (4) What is the ionization rate for the cold ISM? \\ \\ (5) What are the physical conditions of the cold, atomic gas in other galaxies and how do these compare to other properties like jets, winds, and stellar, molecular \& dust mass?}\\

Instead of attempting to summarize here all of the possible science topics that can be pursued with CRRLs and the many interesting regions that can be investigated we have chosen to limit our discussion in this section to two prime targets in the southern hemisphere, the Galactic center and the Magellanic clouds. In the Sect.'s 5 and 6 we will then discuss the technical details for studying these objects with SKA-LOW in phase 1 and phase 2.\\

\noindent \textbf{SKA \& Galactic Center}. The center of the Milky Way shares many properties with other, often more spectacular nuclei of galaxies. Indeed, at a distance of only 8 kpc, the nucleus of our galaxy is a unique laboratory for studies of global phenomena -- e.g., accretion, mass flow, star formation -- and the physical and chemical conditions and the micro-physics processes at work in the interstellar medium in the nuclei of galaxies. The few parsecs surrounding the massive black hole in the center of our Milky Way contain a dense and luminous star cluster, as well as various components of neutral, ionized and hot gas. 

The central parsec region contains mostly ionized gas (the HII region Sgr A West) and hot, X-ray emitting gas. This low density, ionized 'central cavity' is pervaded by a set of orbiting ionized gas filaments, which in turn are surrounded by orbiting, dense molecular cloud streamers at $\sim$1.5 to 4 pc (the circum-nuclear disk). The outer edge of the circum nuclear disk is bordered by the young supernova remnant, Sgr A East, and surrounded by a number of dense molecular clouds on a scale of 5 to 100 pc (see the reviews by Genzel et al. 2010; Morris \& Serabyn 1996, and references therein). This central molecular zone is embedded in a more extensive HI nuclear disk.  The high surface density and total mass content likely reflect inflow from larger radii.

Because of the high pressures and strong tidal shearing effects, the average properties of the molecular clouds in the Galactic Center differ significantly from those in the outer Galaxy with elevated gas temperatures, high average densities and strong turbulent velocity fields. Magnetic fields with mG field strength have been reported, that -- at least locally -- control the dynamics of the gas. A number of gas heating mechanisms are important in this nuclear environment including far-UV radiation emitted by the OB stars near the center responsible for the HII regions and the surrounding photo-dissociation regions at the edges of the cavities blown by the hot stars. Clouds compressed at the edges of the expanding bubbles are shock heated. The dense gas phase in shielded cloud cores is heated by turbulent dissipation and magnetic viscous heating. The interrelationship of these structures and the processes that drive them will be an important testbed for our general understanding of what drives the evolution of galactic nuclei. 

Low-frequency CRRLs can provide a unique information on the central region of our galaxy specifically on the physical conditions in and the dynamics of the dense photodissociation regions that separate the ionized and molecular gas (Wyrowski et al. 1997, 2000; Natta et al. 1994). On a larger scale, the carbon radio recombination lines can probe the interrelationship of the molecular gas and the wider atomic H layer.\\

\noindent \textbf{The Magellanic Clouds}. The Large Magellanic Cloud (LMC) and Small Magellanic Cloud (SMC) are irregular galaxies in orbits around the Milky Way that have experienced several tidal encounters with each other and with the Galaxy. The Large Magellanic Cloud contains a bright stellar bar which together with regions of star formation and the patchy dust obscuration dominates its optical appearance. In HI, on scales of 15 to 500 pc, the morphology is dominated by HI filaments with numerous shells and holes. The turbulent and fractal nature of the ISM on these scales is the result of dynamical feedback into the ISM from star formation processes. On larger scales, (200 pc to 10 kpc), the HI distribution is very patchy. The reduced metallicity implies lower dust shielding and higher photodissociation rates. Hence, the Magellanic Clouds are a more hostile environment for molecular gas. Indeed, while -- because of self-shielding -- much of the hydrogen may be in molecular form in the clouds, carbon may be largely in the form of C$^{+}$. In general, this so-called CO-dark gas may be an important reservoir of gas in low metallicity environments.   

The LMC is home to the most luminous and best known massive star-forming region in the Local Group, the Tarantula nebula, powered by the super star cluster 30 Dor, which contains some 1000 bright O stars. The Magellanic Clouds serve as excellent laboratories for the study of the process of star formation in an environment that, in many ways (such as metallicity, dust content, and star formation rate), resembles the extreme conditions of the early universe and distant star-forming regions. 

The LMC and SMC are also excellent astrophysical laboratories for studying the life cycle of the interstellar medium (ISM). Located at 50 and 60 kpc, their proximity permits detailed studies of resolved ISM clouds and their relation to stellar populations on global scales, which in the Milky Way are obfuscated by line-of-sight crowding and distance ambiguity. The SMC and LMC (and Milky Way) span an interesting range in their global properties (metallicity, molecular gas content, and gas-to-dust ratio) that make them interesting testbeds for galaxy evolution studies.  In terms of galaxy evolution, the LMC--SMC pair is well suited for studying how the agents of evolution, the ISM and stars, operate as a whole in two galaxies that are tidally interacting with each other and with the Milky Way. 

Low frequency carbon recombination lines will provide a unique tool specifically geared towards tracing CO-dark gas and its relationship to, on one hand, HI clouds and molecular gas and, on the other hand, the regions of active star formation. In addition, CRRLs can be used to study the physical conditions in the photodissociation regions associated with regions of massive star formation. In particular, the 30 Dor region will allow a study of the characteristics of PDRs in a super star cluster environment. Finally, we like to highlight that the many bright HII regions and supernova remnants will provide convenient bright background sources for detailed studies of the carbon and hydrogen recombinations that can probe the cosmic ray ionization rate on a galaxy-wide scale and relate them to their likely sources. 

\section{SKA Phase 1}
The technical details of SKA-LOW in phase 1, SKA1-LOW, are described in the SKA1 Imaging Science Performance document (hereafter SKA1-ISP) by R. Braun. SKA1-LOW provides an increase in sensitivity, relative to its precursors and pathfinders, by up to 2 orders of magnitude. The proposed configuration offers a dense core, with a filling factor close to unity. Outside the core the stations are arranged in three outwards extending spiral arms. The core dominates the angular resolution versus sensitivity and the sweetspot is in range 1-5 arcmin at 140 MHz. SKA1-LOW will have a large bandwidth from 50-350 MHz where up to 243 $\alpha$-transition CRRLs (n=266-508) can be observed simultaneously. Comparing with the models in Fig.~2 shows that this range is well suited to study the low density, low temperature CRRL emitting gas.\\

\noindent \textbf{Galactic survey}. The overall goal of a Galactic CRRL survey with SKA1-LOW is to determine the physical conditions of the cold, diffuse clouds and their role in the process of starformation in the inner Galactic plane (l$<$50 deg, b$<$|5| deg). In Sect. 2 we showed that, at frequencies below 350~MHz, on degree-scales Galactic CRRLs have typical line widths in the range from 10 to 50 of km/s and typical peak optical depths in the range of a few times 10$^{-4}$ to a few times 10$^{-3}$.

Table~\ref{t1} shows that upon stacking the CRRLs in sets of 9 lines that the SKA1-LOW sensitivity in 8~hours with 1~kHz resolution is sufficient to detect CRRLs down to a peak optical depth level of $\sim$10$^{-4}$ throughout the entire inner Galactic plane on scales from 1.5 arcmin at 350 MHz to 10 arcmin at 50 MHz. The arcmin-scale resolution is well suited to study the cold, diffuse clouds in the Milky Way and the peak optical depth limit corresponds to a HI column density of about 1$\times$10$^{20}$~cm$^{-2}$.

Stacking CRRLs in sets of 9 does not affect our science goals as the lines are spaced very closely. In this way we obtain 27 stacked CRRL $\alpha$ transition measurements for each line of sight across the full frequency range. In Fig.~2 we give an example of what such a measurement with SKA1-LOW simulated noise could look like. The poorest sensitivity, in units of optical depth, is obtained at 350 MHz. However, by convolving the higher frequency maps to the spatial resolution of the lowest frequency maps at 50~MHz the noise will be similar at all frequencies. By comparing the SKA1-LOW noise levels with our models this means that we can determine the electron temperature and density on arcmin-scales in diffuse clouds to better than a factor 2 down to a limiting HI column density of about 1$\times$10$^{20}$~cm$^{-2}$.

Additional constraints on the electron temperature and density of the cold gas can be obtained from the CRRL line width (e.g. Kantharia et al. 1998). However, the currently planned 1 kHz channel width in phase 1 is not sufficient to resolve the narrowest CRRLs measured towards Cas~A (e.g. Fig.~3; Payne et al. 1989) and Cygnus~A (Oonk et al. 2014). To resolve these lines would require a channel width of about 0.3 kHz.

A 500 deg$^{2}$ survey of the inner Galactic plane requires 13 pointings at 50 MHz and 500 pointings at 350 MHz. For 8~hr pointings this would take about 10$^{2}$~hrs at 50~MHz and 4$\times$10$^{3}$ hrs at 350~MHz and is therefore limited by the small field of view at 350 MHz. If a beamformer becomes part of the SKA1-LOW design this time requirement can be reduced by scanning the sky in smaller frequency chunks with more beams.

One of the prime targets of a Galactic CRRL survey with SKA1-LOW is the Galactic center. At the distance of the Galactic center (8 kpc) the 1.5 arcmin resolution at 350~MHz means that we can resolve CRRL emitting structures down to 4~pc at 350~MHz. To study the physical properties of these structures we are limited to the 10 arcmin resolution at 50~MHz corresponding to about 25 pc. This means that in phase 1 we can study the interaction of the molecular gas with the wider atomic H layers. Smaller scale structures such as the PDRs and the HI nuclear disk require the capabilities of SKA-LOW in phase 2.

To derive the carbon abundance and convert our carbon column densities to hydrogen column densities we aim to combine our CRRL measurements with the planned Galactic HI absorption grid survey, see McClure-Griffiths et al. (2014) in this volume. We estimate that this can be done on average in the Galactic plane for about 5 extragalactic background sources per deg$^{2}$ (e.g. Wilman et al. 2008; Riley et al. 1999). This number should be considered a lower limit for measuring the carbon abundance in the galactic center given the many bright Galactic radio sources that are located in this region and which can also be used for this purpose.

The models in Fig.~2 and the Cas~A spectra in Fig.~3 show that with with the spatial resolution and sensitivity of SKA1-LOW it should be possible to detect the narrow HRRL component from diffuse clouds and separate it from the broader HRRL components associated with the WIM. The models show that this is best done at higher frequency range of SKA1-LOW. In combination with the corresponding CRRLs this will allow us to determine the hydrogen ionization rate due to cosmic rays in these clouds. Presently, this rate in the CNM has only been measured through molecular observations in the infrared or submillimeter (e.g. Indriolo et al. 2007; Shaw et al. 2008). However, these observations have been limited to a small number of bright sources and the analysis is hampered by the complex molecular physics involved. Cosmic ray ionization, especially in the Galactic center, may play an important role in heating of diffuse clouds.\\

\noindent \textbf{Extragalactic survey}. The overall goal of an extragalactic CRRL survey with SKA1-LOW is to determine the presence of CRRLs in other galaxies and use these lines to trace the physical conditions of the CNM in these objects. Our CRRL models show that we can expect a peak optical depth of about $10^{-3}$ for a HI column density of about 10$^{21}$~cm$^{-2}$ (Salgado et al. in prep.). This is consistent with our recent detection of CRRLs in M82 (Morabito et al. 2014). Furthermore the known frequency spacing of RRLs can be used to determine redshifts for extragalactic objects. This is particularly useful for those optically faint objects where it is difficult to identify the HI 21~cm line. 

Table~\ref{t1} shows that upon stacking the CRRLs in sets of 9 lines that the limiting point source sensitivity is about 0.4~Jy at 160 MHz for an 8~hr observation with SKA1-LOW and a resolution of about 3 arcmin. A 2$\pi$~sr survey to this depth would enable a CRRL search in up to 10$^{5}$ galaxies, spanning a redshift range from z=0 to 5 (Wilman et al. 2008). This sample size would represent an increase of almost 3 orders magnitude relative to our planned LOFAR survey.

The time requirement for such a 2$\pi$~sr survey varies between 4$\times$10$^{3}$ hrs at 50 MHz and a staggering 1.6$\times$10$^{5}$ hrs at 350 MHz. At the high frequencies this is clearly too long to be realistic. If a beamformer becomes part of the SKA1-LOW design then these time requirements can be reduced significantly. Alternatively, one could first perform a blind CRRL detection survey near 100~MHz and then only follow-up the detected galaxies with wider frequency coverage using pointed observations.

Of particular interest in an extragalactic SKA1-LOW survey are nearby galaxies. We estimate that for 8~hr integrations about 400 nearby galaxies are bright enough to be detected CRRLs and that in about 100 of these the CRRLs can be spatially resolved (e.g. Israel 1990, Negrello et al. 2013). A pointed survey for nearby galaxies over the full frequency range would therefore take about 4$\times$10$^{3}$ hrs.

Two prime targets in a nearby galaxy survey are the Magellanic Clouds. They contain large quantities of cold, atomic gas (e.g. Stanimirovic et al. 1999; Dickey et al. 2000) and have a 408~MHz background radio continuum temperature of about 30-120 K with an approximate $\lambda^{2.6}$ scaling to lower frequencies (Haslam et al. 1982; Israel et al. 2010). Table~\ref{t1} then shows that in an 8~hr single pointing we can reach peak optical depths of few times 10$^{-3}$ at 350 MHz to a few times 10$^{-4}$ at 60~MHz. This is sufficient to detect CRRLs for hydrogen column densities down a few times 10$^{20}$~cm$^{-2}$ on 10 arcmin scales.  

At the distance of the Magellanic Clouds a 10 arcmin scale corresponds to 200 pc. This means that we will be able to resolve the CRRLs associated with larger scale patchy HI distribution and HI supershells. The smaller scale HI shells and PDRs require the capabilities of SKA-LOW in phase 2. Just as for the inner Galactic plane survey we estimate there are on average 5 extragalactic background sources per deg$^{2}$ for which HI 21~cm absorption and CRRL measurements can be compared to derive the carbon abundance in these systems.

\subsection{SKA Phase 1: Early science}
For early science (ES) it is expected that SKA1-LOW will have about half of its final sensitivity. The CRRL science that can be done during ES depends the array configuration. If the filling factor of core of the array during ES remains similar to that of the final array configuration then the inner Galactic plane survey, as outlined above, can be carried out. Given the high HI column densities and high background continuum radiation the survey can then start at the Galactic center during ES and progress outwards in Galactic longitude as the array builds up to its full phase 1 sensitivity.

For an extragalactic CRRL survey the lower sensitivity during ES has different effects for different source populations. For the bright, distant FR~I and FR~II radio galaxies the number of objects that can be searched for CRRLs down to a peak optical depth of 10$^{-3}$ decreases with about a factor of two (Wilman et al. 2008). For nearby galaxies the effect is more severe. We estimate a decrease by about a factor 4, such that at most 100 nearby galaxies are bright enough to be searched for CRRLs and that only about 20 of these are sufficiently bright and large that the CRRLs within can be spatially resolved (Negrello et al. 2013, Israel et al. 1990).

For the Magellanic Clouds the lower sensitivity during early science will primarily affect the limiting column density. Therefore for a fixed observing time of 8~hrs per pointing we will start to miss the more diffuse clouds starting at a HI column density of about 5$\times$10$^{20}$~cm$^{-2}$.

\section{SKA Phase 2}
In phase 2 SKA-LOW is expected to have 4 times the sensitivity of SKA1-LOW over a wide range of angular scales and up to 20 times the angular resolution. For both a Galactic and an extragalactic survey this increase in resolution is very important. In the Galactic center we will be able to spatially resolve the HI nuclear disk and connect its kinematical and thermal properties to that of wider inflow of cold gas. Similar this increase in resolution is essential to spatially resolve PDRs and HI shells, as tracers of stellar feedback, over a wide range of environments in both our own Galaxy and the Magellanic Clouds.

For a larger extragalactic survey the increased sensitivity and resolution of SKA-LOW in phase 2 would allow us spatially resolve CRRLs in up to 3$\times$10$^{3}$ nearby galaxies and search for CRRLs in up to half a million $>$0.1~Jy radio sources spanning a range in redshift from z=0 to 5. This will provide important input on how the physical properties of the CNM evolve with time and environment. If redshift information is available then the average CNM properties for well-defined samples of even fainter radio sources can be studied by employing stacking techniques.

\section{Summary}
SKA observations of low-frequency CRRLs, in combination with HI~21~cm absorption measurements, will provide a comprehensive and unprecedented inventory of the low-density, cold ISM in our Galaxy, the Magellanic Clouds and beyond. This will enable us to quantify its role in the overall pressure, mass and energy balance of the interstellar medium and how it relates to the cycles of star formation and death. 

In addition, for studies of the Epoch of Reionization and the HI dark ages it is important to quantify the distribution of the Milky Way CNM, in terms of strength and velocity, as a potential foreground signal. Deep, high spectral resolution, low-frequency RRL observations with SKA-LOW can help provide this quantification and by doing this with the same instrument at the same frequencies one can minimize the uncertainties due to instrumental systematics and extrapolation from higher to lower frequencies.    

In order to spectrally resolve CRRLs associated with the coldest, atomic gas structures on small scales a channel width of about 0.3~kHz will be required, instead of the 1~kHz that is currently planned in the SKA1 design. Another important element that could significantly speed up large low-frequency surveys is a beamformer. The change in field of view with frequency for SKA-LOW means these surveys can be done much more efficiently by trading bandwidth against beams.

\bibliographystyle{apj}

\newpage

%
%
%

\begin{table*}[h!]
\begin{tabular}{l r r r r}
\hline
\hline
 Frequency &  $\sigma_{T,rms,extended}$ & $\tau_{T,peak,extended}$ & $\sigma_{\nu,rms,point}$ & F$_{\nu,point}$  \\ \hline
 [MHz]     &  [mK]     &   $\times$(10$^{-4}$) & [mJy] & [Jy] \\ \hline
 60        &  980      &   (0.5-0.05)         & 1.9 & 1.9   \\
 160       &  140      &   (1.0-0.10)         & 0.4 & 0.4   \\
 220       &  140      &   (2.1-0.21)         & 0.2 & 0.2   \\
 350       &  154      &   (7.5-0.75)         & 0.2 & 0.2   \\ 
\hline
\end{tabular}
\caption[]{CRRL sensitivity with SKA1-LOW. Column (1) Observing frequency. Column (2) Core SKA1-LOW 5$\sigma$ spectral sensitivity for extended sources, in units of mK, for an effective baseline B$_{eff}$=2~km with 1~kHz channels, 8~hr integration and stacking 9 lines. Column (3) Corresponding 5$\sigma$ peak optical depth range for extended sources. These ranges are calculated for the inner Galactic plane with continuum temperatures between 140 and 1400 K at 408~MHz (Haslam et al. 1982) and using a $\lambda^{2.5}$ scaling (e.g. Rogers et al. 2008). Column (4) Full SKA1-LOW 5$\sigma$ point source spectral sensitivity, in units of flux, for 30~km/s channel width, 8~hr integration, 200 arcsec FWHM beam and stacking 9 lines. Column (5) Corresponding limiting point source continuum flux F$_{\nu,lim}$ to obtain a 5$\sigma$ CRRL detection with $\tau_{peak}$=10$^{-3}$. Comparing to the S3 simulation (Wilman et al. 2008) and the 7C survey (Riley et al. 1999) shows that on average there are about 5 extragalactic sources per deg$^{2}$ brighter than these limiting fluxes. For both the extended and point source cases SKA1-LOW is typically about a factor 10 more sensitive than LOFAR at 160 MHz and a factor 100 more sensitive than LOFAR at 60 MHz. The above sensitivities are derived from the SKA-ISP document.}\label{t1}
\end{table*}

%
%
%

\begin{figure*}[h!]
\mbox{
 \includegraphics[width=0.36\textwidth, angle=90]{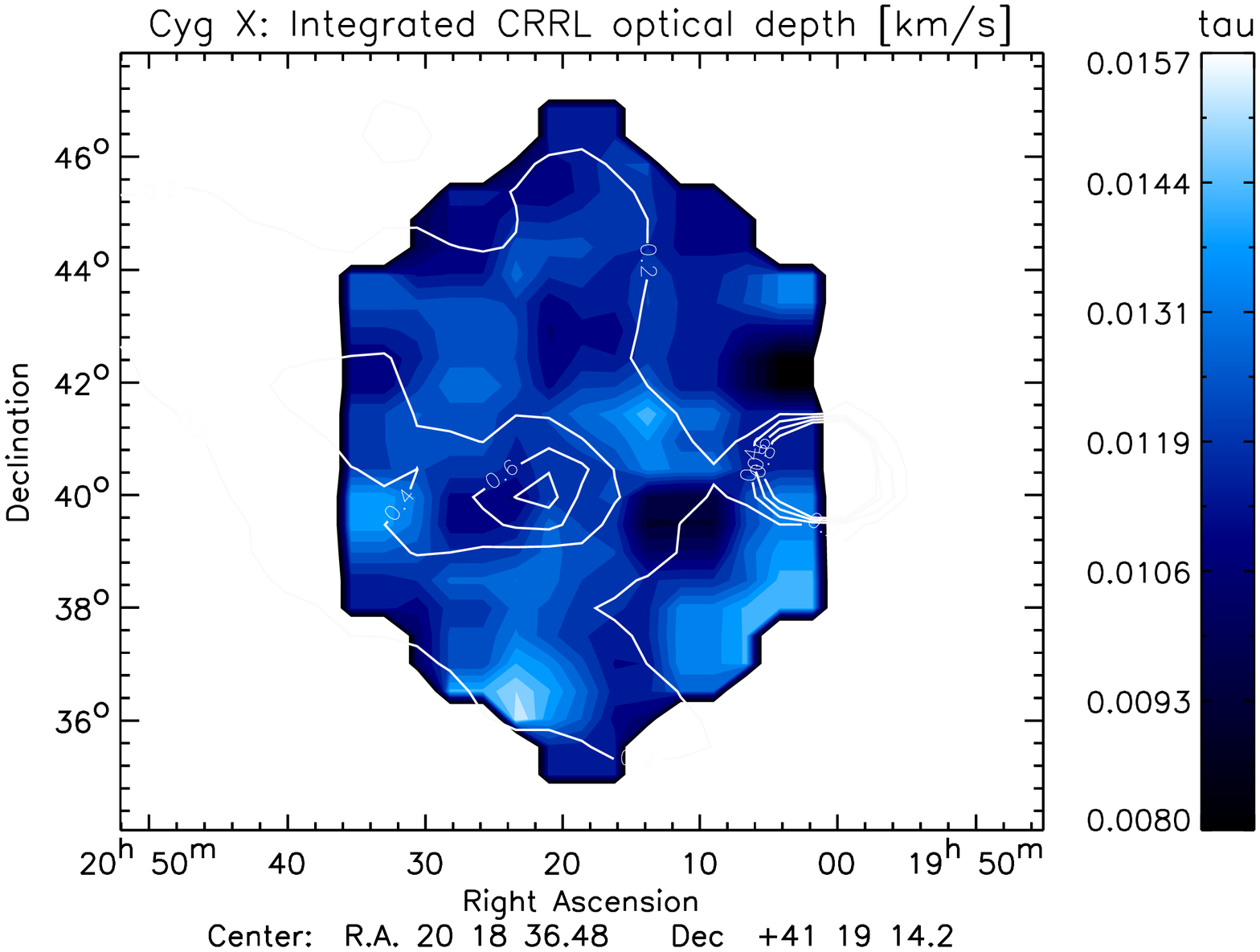}
 \hspace{0.2cm}
 \includegraphics[width=0.36\textwidth, angle=90]{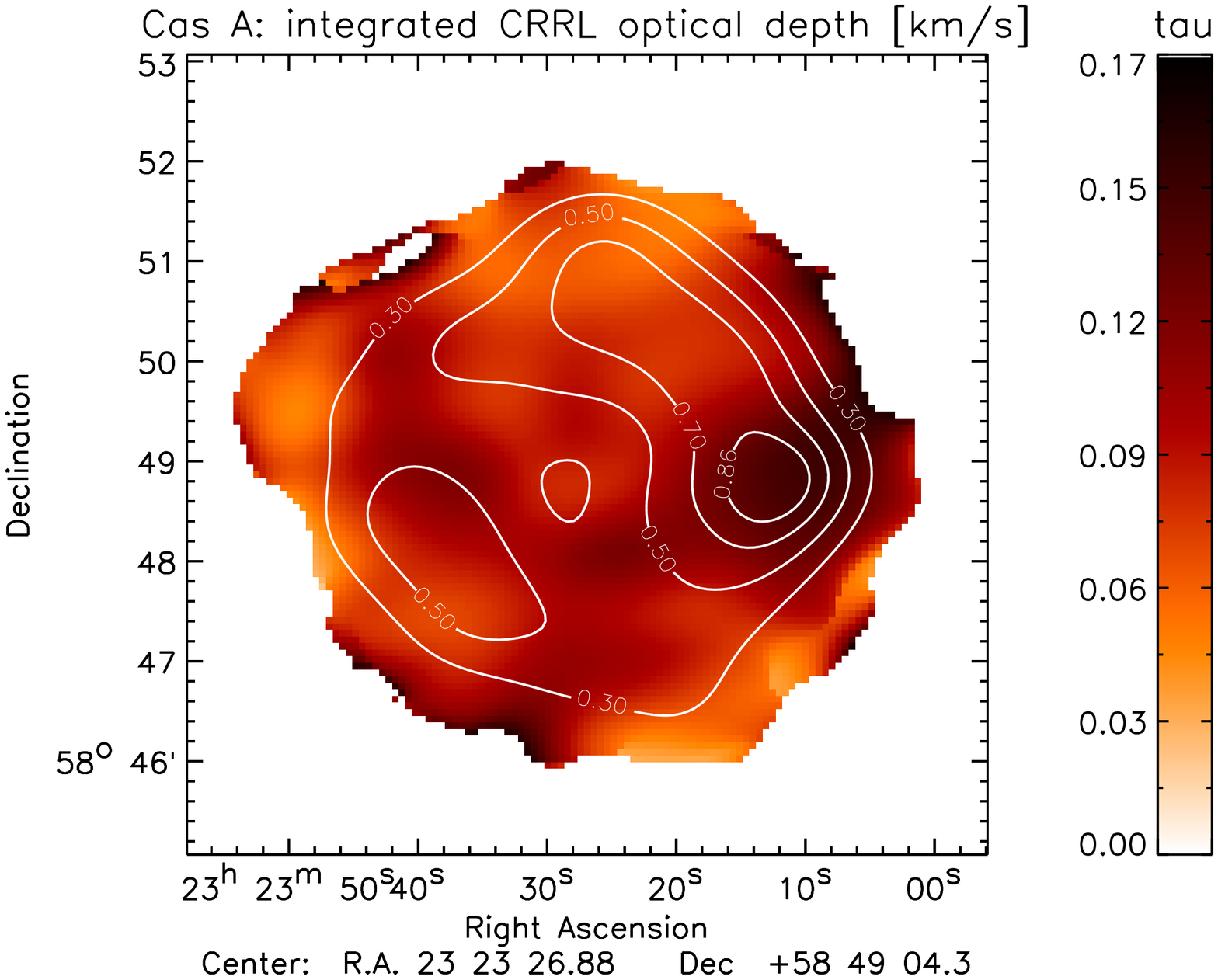}
}
\vspace{-0.5cm}
\caption[]{Observations of CRRL absorption in the Milky Way near 50 MHz (Oonk et al. in prep.). \textit{(Left)} LOFAR 4~hr observation of degree-scale CRRLs in the Cygnus~X region. Shown is a map of the integrated CRRL absorption in units of optical depth with radio continuum contours overlaid in white. The map shows structure down to the 1 degree limit of this observation. \textit{(Right)} LOFAR 8~hr observation of arcmin-scale CRRLs along the line of sight to Cas~A. Shown is a map of the integrated CRRL absorption in units of optical depth with radio continuum contours overlaid in white. The CRRL absorption is resolved in several filamentary structures that agree well with the corresponding HI 21~cm absorption (e.g. Bieging et al. 1991).}\label{fmwbf}
\end{figure*}

\begin{figure*}[h!]
\center
\mbox{
 \hspace{-0.6cm}
 \includegraphics[width=0.37\textwidth, angle=90]{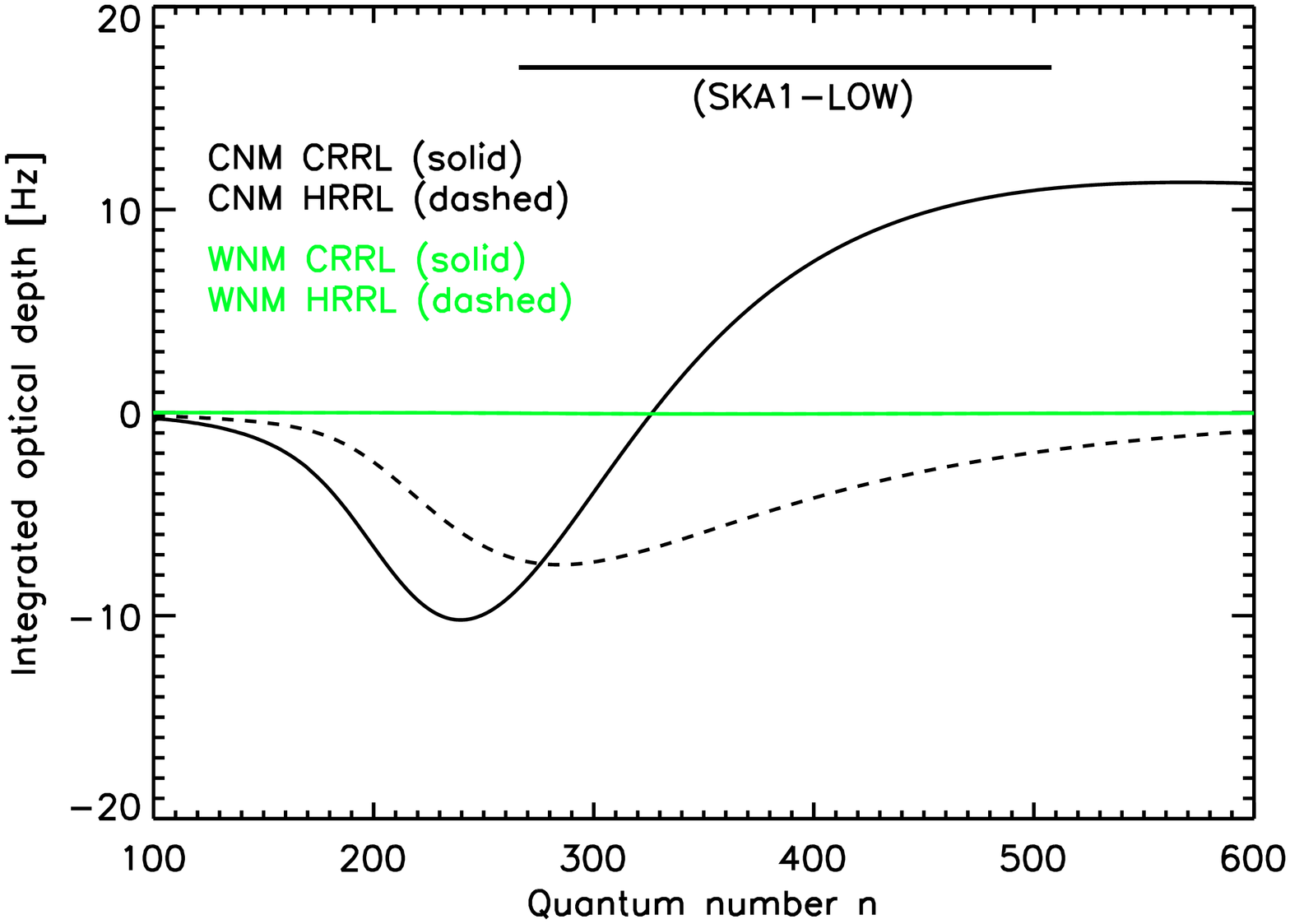}
 \hspace{0.2cm}
 \includegraphics[width=0.37\textwidth, angle=90]{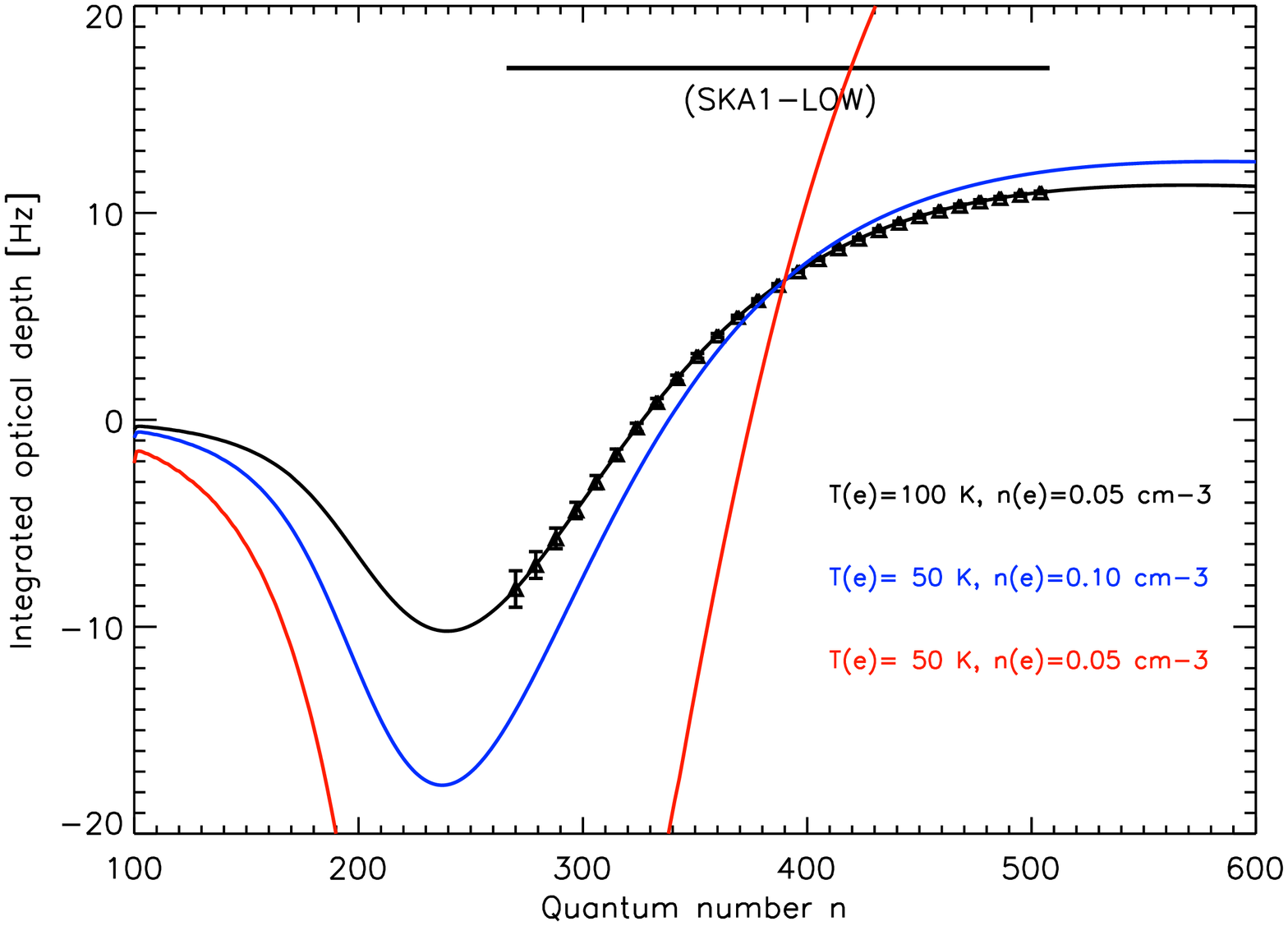}
}
\vspace{-0.5cm}
\caption[]{RRLs and the diffuse neutral ISM. Model results are given for the CNM (n$_{H}$=50~cm$^{-3}$, n$_{e}$=0.05~cm$^{-3}$, T$_{e}$=100~K) and WNM (n$_{H}$=0.1~cm$^{-3}$, n$_{e}$=0.01~cm$^{-3}$, T$_{e}$=10000~K). For both the CNM and the WNM a hydrogen column density N(HI)=10$^{21}$~cm$^{-2}$ is used.\textit{(Left)} Integrated optical depth vs. quantum number for CRRL (solid) and HRRL (dashed) for the CNM (black) and WNM (green). The WNM optical depth is negligible and as such the HRRL and CRRL curves for the WNM overlap on this plot. \textit{(Right)} The CRRL curve for the CNM model with error bars simulated for a 8~hr single pointing core SKA1-LOW observation in the inner Galactic plane with a continuum temperature of 140~K at 408 MHz. The integrated optical depth depends linearly on the column density. In red and blue we overplot two lower temperature models that are scaled to reproduce the optical depth for the black reference model at n=390.}\label{f1}
\end{figure*}

\begin{figure*}[h!]
\mbox{
 \includegraphics[width=0.37\textwidth, angle=90]{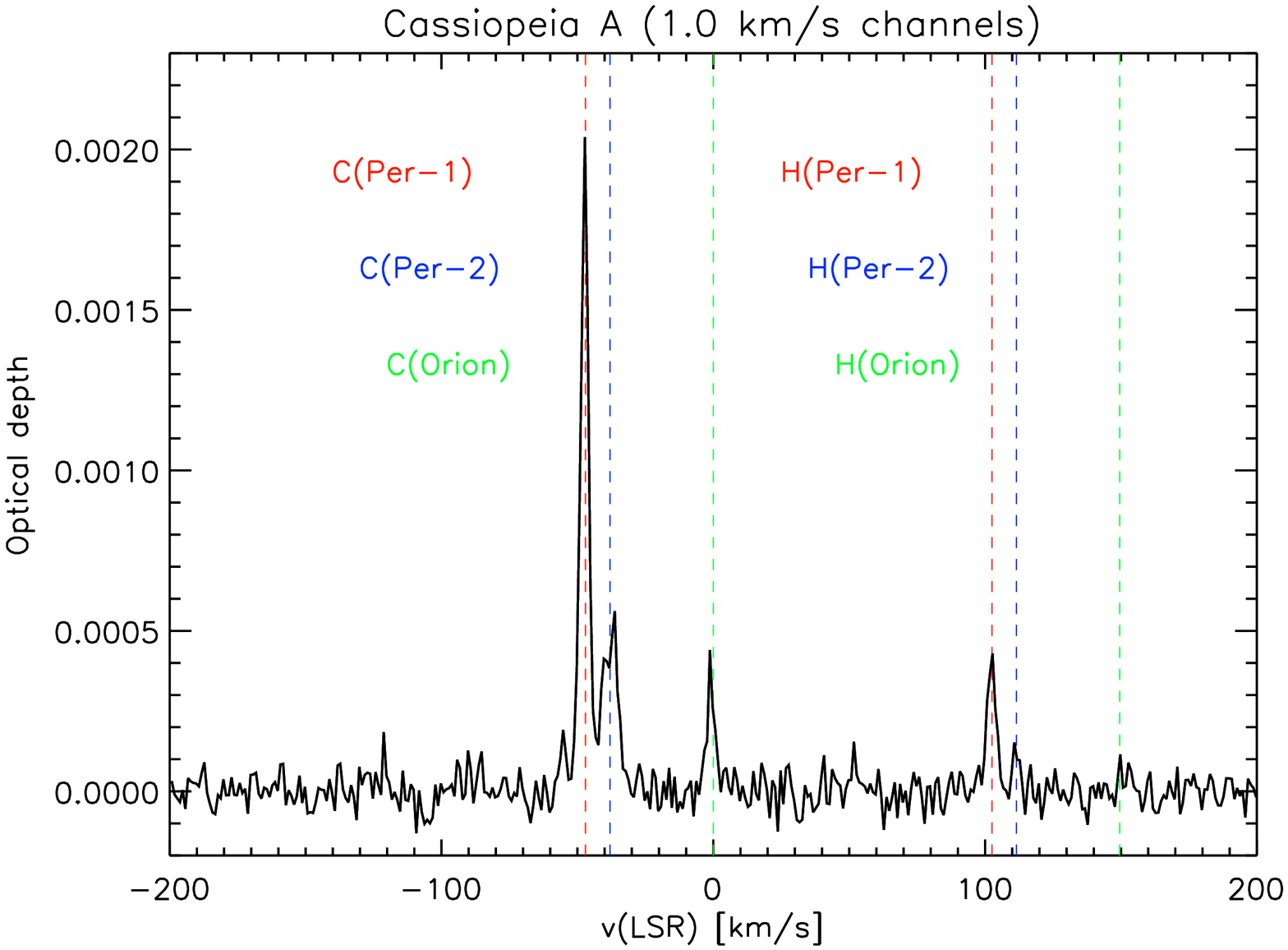}
 \includegraphics[width=0.37\textwidth, angle=90]{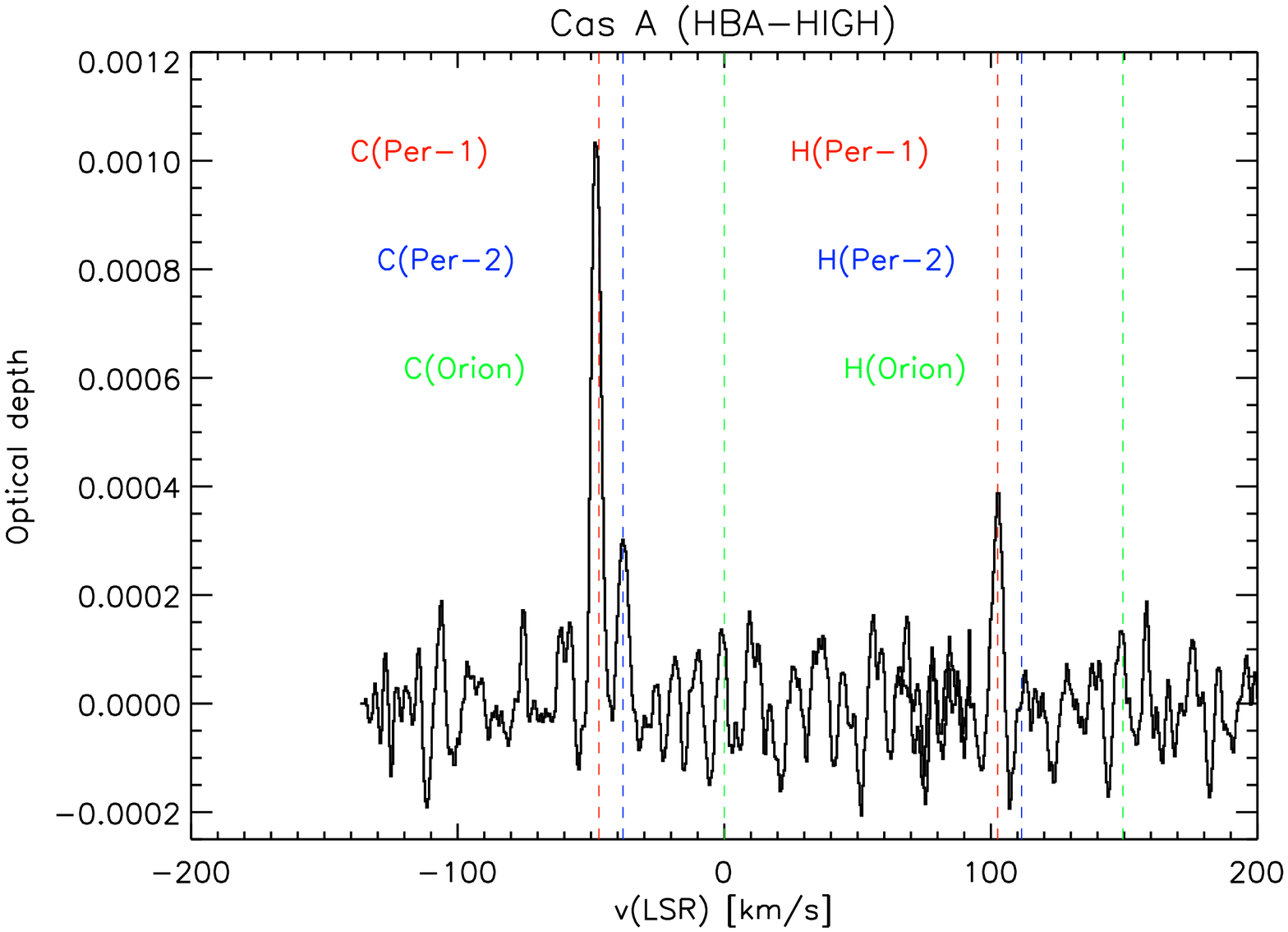}
}
\vspace{0.0cm}
\caption[]{LOFAR and WSRT observations of CRRL and HRRL emission in the cold clouds towards Cas~A. \textit{(Left)} WSRT 12~hr 350 MHz observation. The RRL spectrum showing matching narrow HRRL and CRRL emission. For both the WSRT and LOFAR spectrum the velocity scale has been centered on the local standard rest for CRRLs. on this scale the corresponding HRRLs are offset by +150 km/s. \textit{(Right)} LOFAR 3~hr HBA-HIGH observation at 220 MHz. This spectrum was obtained using only the LOFAR core stations. The HRRLs in both the LOFAR and WSRT spectra are likely associated with hydrogen that has been ionized by cosmic rays and the HRRL/CRRL ratio implies a cosmic ray hydrogen ionization rate $\zeta_{H}\sim$10$^{-16}$~s$^{-1}$ (Oonk et al. in prep.).}\label{fcasmap}
\end{figure*}

\end{document}